\documentstyle[11pt,psfig,vsopsymp]{article}
\begin{document}

\markboth{\hfill Snellen et al.}{Space-VLBI observations of 
young radio sources \hfill}

\centerline{\LARGE\bf 
   A study of young radio-loud AGN 
}\medskip
\centerline{\LARGE\bf using space-VLBI}\bigskip
\renewcommand{\thefootnote}{\fnsymbol{footnote}}
\centerline{\sc 
Ignas Snellen$^1$
\footnote{This research was supported by the European Commission, 
TMR Programme, Research Network Contract ERBFMRXCT96-0034 ``CERES'', and
the TMR Access to Large-scale Facilities programme under contract No. 
ERBFMGECT950012}, 
Wolfgang Tschager$^2$, Richard Schilizzi$^{3,2}$
}\medskip
\centerline{\sc 
Huub R\"ottgering$^2$ \& George Miley$^2$
}\medskip

\centerline{\it
$^1$ Institute of Astronomy, Madingley Road, Cambridge CB3 0HA, UK}
\centerline{\it
$^2$ Sterrewacht Leiden, Postbus 9513, 2300 RA Leiden, The Netherlands}

\centerline{\it
$^3$ Joint Institute for VLBI in Europe, Postbus 2, 7990 AA, Dwingeloo,}
\centerline{\it The Netherlands}

\begin{abstract} \noindent 
Gigahertz Peaked Spectrum (GPS) sources form a key element in the 
study of the onset and evolution of radio-loud AGN, since they are most likely 
the young counterparts of extended radio sources.
Here we discuss space-VLBI observations of GPS sources, which enable us 
to obtain unprecedented angular resolution at frequencies near their
spectral turnovers. Observed peak brightness temperatures of $10^{10.5-11}$
Kelvin indicate that synchrotron self absorption is responsible for 
their spectral turnovers. This is in close agreement with previous size $-$ 
spectral turnover statistics for GPS sources. The combination of these
new space-VLBI observations with ground-based VLBI observations taken at an
earlier epoch, confirm the young ages for the 
most compact GPS galaxies of several hundred years.
\end{abstract}


\sources{0108+388, 2021+614}

\section{Young radio-loud AGN}

Although radio-loud Active Galactic Nuclei (AGN) have been studied
for several decades, still not much is known about their birth
and subsequent evolution. The recent identification of a class of 
very young radio sources can be considered as a major breakthrough
in this respect, since it has opened many unique opportunities for 
radio source evolution studies.

Unfortunately, the nomenclature and use of acronyms in this 
field of research is rather confusing. This is mainly caused by
the different ways in which young radio sources are selected.
Selection of young sources is made in two ways, the first based  
on their broadband radio spectra, and the second based on their
compact morphology. 
A convex shaped spectrum, peaking at about 1 GHz 
distinguishes young radio sources from other classes of compact radio sources.
In this case they are called Gigahertz Peaked Spectrum ({\bf GPS}) radio
sources (eg. O'Dea etal. 1991, O'Dea 1998). Similar objects, which are
typically an order of magnitude larger in size, have their spectral turnovers 
shifted to the $10 - 100$ MHz regime, causing them to be dominated  at cm 
wavelengths by the optically thin parts of their spectra. 
These are called Compact Steep Spectrum ({\bf CSS}) radio
sources to distinguish them from the general population of extended 
steep spectrum sources (eg. Fanti et al. 1991). 

On the other hand, young radio sources are found in multi-frequency VLBI
surveys, in which they can be recognised by  compact jet/lobe-like
structures on both sides of their central core. They are called 
Compact Symmetric Objects ({\bf CSO}, Wilkinson et al. 1994). 
Their double sided structures
clearly distinguish them from the large majority of compact sources
showing one-sided core-jet morphologies. This implies that the luminosities
of CSO are unlikely to be substantially enhanced by Doppler boosting. 
Larger versions
of CSOs are subsequently called Medium Symmetric Objects ({\bf MSO}) and 
Large Symmetric Objects ({\bf LSO}).

The overlap between the classes of CSO and GPS galaxies is large and we 
believe that they can be considered to be identical objects.
However, note that a substantial fraction of GPS sources are optically 
identified with high redshift quasars, which in general show core-jet 
structures (Stanghellini et al. 1997). 
The relationship between GPS quasars and GPS galaxies/CSO
is not clear and under debate (Snellen et al. 1999). 
We therefore believe it is wise to
restrict evolution studies to GPS galaxies and CSOs.

\subsection*{Evidence for youth}

Although it was always speculated that GPS sources were young objects,
only recently has strong evidence been found to support this 
hypothesis. Monitoring several GPS sources over a decade or more using
VLBI, allowed Owsianik \& Conway (1998) and Owsianik, Conway \& 
Polatidis (1998) to 
measure the hotspot advance speeds of several 
prototype GPS sources to be $\sim 0.1h^{-1}c$. These imply dynamical 
ages of typically $10^{2-3}$ years. 

Additional proof for youth comes from analysis of the overall radio
spectra of the somewhat larger CSS sources. Murgia et al (1999) show that 
their spectra can be fitted with synchrotron aging models, implying 
ages of typically $10^{3-5}$ years.

The work of these authors shows that GPS/CSO sources are very young and most 
likely 
the progenitors of large, extended radio sources. This makes them
key objects for radio source evolution studies.

\subsection*{Tools for radio source evolution studies}

Several authors have used number count statistics and 
linear size distributions to constrain the luminosity evolution
of radio sources (Fanti et al. 1995; Readhead et al. 1996, O'Dea \& 
Baum 1997). All these studies find an excess of young 
objects in relation to the number of old, extended radio sources.
This over-abundance of GPS and CSS sources has generally been explained by
assuming that a radio source significantly decreases in luminosity over its 
lifetime. In this way, sources are more likely to 
contribute to flux density limited samples at young than at old age, causing 
the apparent excess.

However, in addition to their over-abundance, GPS galaxies are found to be 
significantly 
more biased towards high redshift than large extended radio galaxies (Snellen \& Schilizzi, 2000). 
This is puzzling since classes of sources representing similar objects at 
different stages of their evolution are expected to have similar birth 
functions and redshift distributions. Furthermore, it suggests that the 
interpretation of their number count statistics, which are averaged over a 
large redshift range, is not so straightforward.
We have postulated a simple evolution scenario which can resolve these 
puzzles. We argue that the luminosity evolution of a 
radio-loud AGN
during its first $10^5$ years is qualitatively very different from that 
during the rest of its lifetime. This may be caused by a turnover
in the density profile of the interstellar/intergalactic medium 
at the core-radius of the host galaxy, resulting in an increase in
luminosity for young, and a decrease in luminosity for old radio-loud AGN 
with time. Such a luminosity evolution results in a 
flatter collective luminosity function for the young objects, 
causing their bias towards higher 
redshifts, and their over-abundance at bright flux density levels (Snellen et al. 2000).

An alternative explanation is that GPS sources are indeed young AGN, but 
mainly short-lived objects, which will never evolve into extended radio sources
(Readhead et al. 1994). In that case, the two populations are not 
directly connected, and no similar cosmological evolution or redshift 
distribution is necessary.

\section{Space-VLBI observations of GPS sources}

In general, the angular resolution of VLBI observations at a 
certain observing frequency is limited by the size of the earth.
The combination of ground VLBI stations with the Japanese satellite 
HALCA (part of the VLBI Space Observatory Programme 
 VSOP), achieves a resolution typically 3 times higher than this
($\sim 1.5$ mas and $\sim 0.5$ mas at 1.6 and 5 GHz respectively).
In particular, the study of GPS sources benefits from VSOP, since observing at
a higher frequency to achieve a similar resolution is often not an option,
 because of their steep fall-off in flux density towards high frequency.
Furthermore, their physical properties are most interesting around their 
spectral turnover, where differences in spectral indices within the source 
are more prominent than at high frequency.

We have been awarded VSOP observing time for 11 and 8 of the brightest
and most compact GPS sources at 5.0 and 1.6 GHz respectively. 
Details and status of the observations are listed in Table \ref{table1}.
At the time of writing, all targets at 5 GHz,  and 
6 of the 8 sources at 1.6 GHz have been observed.

\begin{table}
\caption{Status of VSOP observations. \label{table1} }
\hskip 1cm
\begin{tabular}{|lllcc|} \hline
Name & id & z & Date of Obs & Date of Obs\\
     &    &   &     5 GHz   &    1.6 GHz \\ \hline
0108+388&Gal& 0.669&1999.08.06 &1999.08.05 \\
0248+430&QSO& 1.316&1999.02.15 &1999.08.18 \\
0552+398&QSO& 2.370&1999.03.23 &1999.01.15 \\
0615+820&QSO& 0.710&1999.09.18 &    tbd    \\
0646+600&QSO& 0.460&1999.09.20 &1999.09.27 \\  
1333+459&QSO& 2.450&1998.06.22 &1999.05.28 \\
1404+286&Gal& 0.077&1998.06.30 &   tbd     \\ 
2021+614&Gal& 0.227&1997.11.16 &1999.09.28 \\ \hline
1550+582&QSO& 1.324&1998.07.02 &    --  \\
1622+665&Gal& 0.201&1998.05.24 &    --  \\
0636+680&QSO& 3.180&1999.09.19 &    --  \\ \hline
\end{tabular}
\end{table}

\subsection*{First results and discussion}

\begin{figure}
\vspace{170mm}
\includegraphics{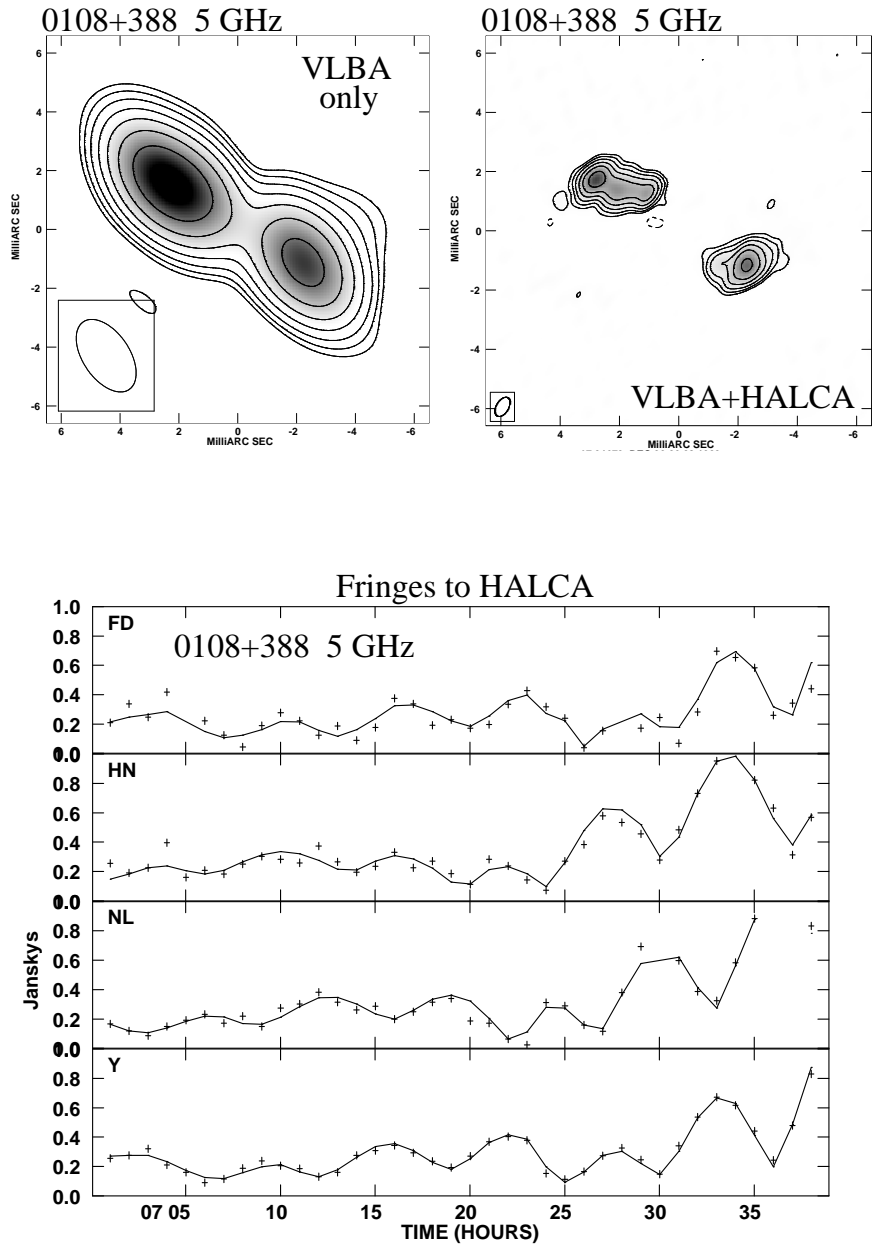}
\caption{\label{fig1} VSOP observations of 0108+388 at 5 GHz. The upper left
panel shows the VLBA only image, the upper right panel shows the VLBA+HALCA
image, and the lower panel shows some of the fringes of the VLBA antennas
and the VLA to HALCA.}
\end{figure}

\begin{figure}
\vspace{80mm}
\includegraphics{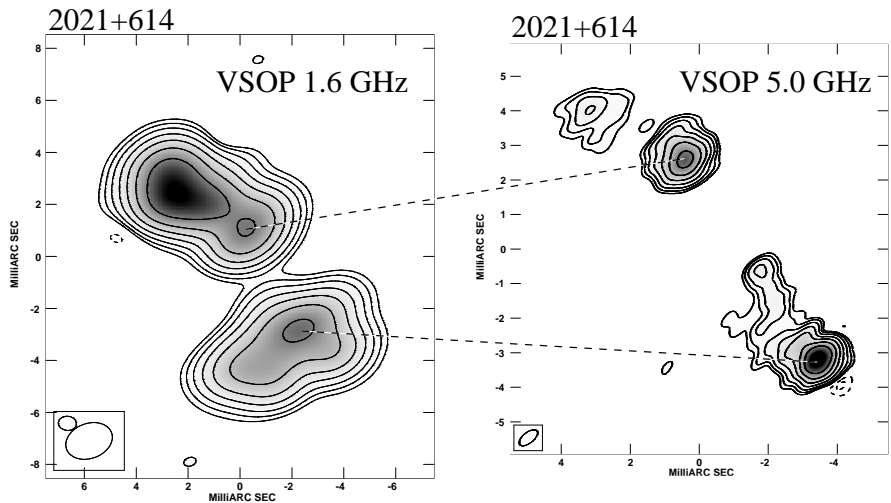}
\caption{\label{fig2} VSOP observations of 2021+614 at 1.6 and 5 GHz. 
The dotted lines connect the two dominant features at 5 GHz with their 
position at 1.6 GHz. Note the importance of sufficient resolution near
the spectral turnover frequency, where the differences in spectral index
between the components are most prominent.} 
\end{figure}

A large fraction of the sources have now been imaged.
Some examples are shown in figures \ref{fig1} and \ref{fig2}.
Additional observations have been taken at 15 GHz with the VLBA to
match the 5 GHz VSOP data in resolution, which will allow detailed  
spectral decompositions of the objects. In particular, this may
sched new light on the nature of the GPS quasars and the role of Doppler 
boosting in these sources.

One of the first results of 
these observations are the high brightness temperatures observed
of typically $10^{10.5-11}$ Kelvin. This indicates that these objects
must be near their synchrotron self absorption (SSA) turnover at the 
observed frequency, making it very likely that indeed SSA is the cause of 
their spectral peaks. This is in agreement with the statistical arguments
of Snellen et al. (2000), who found that among samples of GPS and CSS sources,
the ratio of component size, as derived from the spectral peak assuming SSA,
and overall angular size, are constant and very similar to 
those found for large
extended radio sources. This not only implies self-similar evolution,
but also provides strong 
evidence for SSA. Note however, that several authors argue 
that free-free
absorption can not be ruled out for the smallest GPS galaxies 
(Kameno et al., this volume; Marr et al., this volume)

A valuable spin-off from these high angular resolution VSOP observations
come from their comparison with ground-based VLBI images taken at an 
earlier epoch. Following the method of Owsianik \& Conway (1998), we 
use these to derive dynamical ages for GPS sources.
In this way, we find that the two dominant components at 5 GHz of 2021+614 (fig \ref{fig2}),
have a larger separation at the epoch of the VSOP observations, compared to 
data from Conway et al. (1994) taken in 1982 and 1987. 
The increase in separation indicates
a hotspot advance speed of $\sim 0.1$c, which implies an age of $\sim 400$
years for these components (Tschager et al. 2000). Preliminary analysis 
of 0108+388 (fig \ref{fig1}) shows an advance speed of 15 $\mu as$/yr, 
consistent with what is found by Owsianik, Conway \& Polatidis 
(1998; 9 $\mu as/yr$).
These observations confirm the young ages of a few hundred years for 
the most compact GPS galaxies.

\section{Summary}

GPS galaxies and CSO are now identified as classes of young radio sources.
They form a key element in the investigation of the evolution of 
radio-loud AGN.
We report on VSOP observations of 11 and 8 bright GPS sources
at 5.0 and 1.6 GHz frequency respectively.
First analysis indicates high brightness temperatures consistent with
synchrotron self absorption as the cause of their spectral turnover.
Comparison with ground-based VLBI datasets taken at earlier epochs confirm
the very young ages for the most compact GPS galaxies of a few hundred
years.

\acknowledgements

We gratefully acknowledge the VSOP Project, which is led by the
Japanese Institute of Space and Astronautical Science in cooperation
with many organizations and radio telescopes around the world.

\end{document}